\documentclass[lettersize,journal]{IEEEtran}
\usepackage{amsmath,amsfonts}
\usepackage{algorithmic}
\usepackage[table]{xcolor}
\usepackage{tabularx}
\usepackage{booktabs}
\usepackage{etoolbox}
\usepackage{array}
\usepackage[caption=false,font=normalsize,labelfont=sf,textfont=sf]{subfig}
\usepackage{textcomp}
\usepackage{stfloats}
\usepackage{url}
\usepackage{verbatim}
\usepackage{graphicx}

\usepackage{balance}
\usepackage{nicematrix}
\usepackage[T1]{fontenc}

\definecolor{brown}{RGB}{224,159,82}
\definecolor{pink}{RGB}{254,168,160}
\definecolor{blue}{RGB}{190,224,255}
\definecolor{white}{RGB}{255,255,255}
\makeatletter
\patchcmd{\@makecaption}
  {\scshape}
  {}
  {}
  {}
\makeatletter
\patchcmd{\@makecaption}
  {\\}
  {.\ }
  {}
  {}
\makeatother

\begin{document}
\title{Dynamics of Ideological Biases of Social Media Users}
\author{Mohammed Shahid Modi$^1$, James Flamino$^1$ and Boleslaw K. Szymanski$^{1,2,*}$\\
\vspace{4pt}
\small
\textit{$^1$Department of Computer Science and NEST Center, Rensselaer Polytechnic Institute, Troy, NY, USA}\\
\textit{$^2$Społeczna Akademia Nauk, {\L}\'od\'z, Poland}\\
\textit{$^*$Corresponding Author: szymab@rpi.edu}}

\markboth{IEEE Communications Magazine, May~2024}
{Dynamics of Ideological Biases of Social Media Users}

\maketitle

\begin{abstract}
Humanity for centuries has perfected skills of interpersonal interactions and evolved patterns that enable people to detect lies and deceiving behavior of others in face-to-face settings. Unprecedented growth of people's access to mobile phones and social media raises an important question: How does this new technology influence people's interactions and support the use of traditional patterns? In this article, we answer this question for homophily-driven patterns in social media. In our previous studies, we found that, on a university campus, changes in student opinions were driven by the desire to hold popular opinions. Here, we demonstrate that the evolution of online platform-wide opinion groups is driven by the same desire. We focus on two social media: Twitter and Parler, on which we tracked the political biases of their users. On Parler, an initially stable group of Right-biased users evolved into a permanent Right-leaning echo chamber dominating weaker, transient groups of members with opposing political biases. In contrast, on Twitter, the initial presence of two large opposing bias groups led to the evolution of a bimodal bias distribution, with a high degree of polarization. We capture the movement of users from the initial to final bias groups during the tracking period. We also show that user choices are influenced by side-effects of homophily. Users entering the platform attempt to find a sufficiently large group whose members hold political biases within the range sufficiently close to their own. If successful, they stabilize their biases and become permanent members of the group. Otherwise, they leave the platform. We believe that the dynamics of users' behavior uncovered in this article create a foundation for technical solutions supporting social groups on social media and socially aware networks.
\end{abstract}


\section{Introduction}

\IEEEPARstart{P}{eople} exhibit different patterns of social behavior \cite{Mondani2021} that shape their interpersonal interactions and determine how social groups are created and evolve \cite{dunbar_1993}. Traditionally, these social behaviors have been studied in the context of direct interactions between actors in an offline setting. Hence, their presence and effects within online social environments is not well understood. Indeed, social media has played an ever-growing role in many spheres of human interaction. One such sphere is politics, which is important because it shapes governments and political systems of all levels. In this role, social media provides platforms for politicians to influence countless individuals across vast distances instantly. However, these mediums have also allowed for the widespread dissemination of misinformation \cite{Grinberg2019} and facilitate the polarization of users and the formation of echo chambers (a group of users on social media that exchange information only between themselves, rejecting information from outsiders.) \cite{flamino2021shifting}.

The online interactions in social networks we study here are inherently different from offline face-to-face verbal interactions during which participants silently monitor voice intonation and body language of their partners to recognize their emotions and behavioral patterns. Such recognition facilitates detection of lies and deceiving behavior, but it is missing in online interactions, lowering the chance that social media users will be able to recognize and reject strongly biased, questionable, or faked content. 

We believe these differences create a need for a new understanding of opinion dynamics that is tangential to previous research on human opinion propagation. The DeGroot model \cite{612bb50a-4bdd-3a32-b6eb-7837600cc9c4} describes how an opinion consensus is reached between participants, but isn't designed for users online who can simply switch their opinion group or even drop out as the cost of leaving is much lower than in conventional cases: It is easier to change online groups than face-to-face groups. The Friedkin-Johnsen model \cite{doi:10.1080/0022250X.1990.9990069} handles opinion dynamics in an abstract ``social network'' context, better suited for social media groups. However, it may not account for quitting behavior, the presence of echo chambers, the structural bias of content delivery algorithms and related factors. Accordingly, there is a need to further our understanding of dynamics of social groups in social media to amplify their benefits but temper their drawbacks.

One social principle that is integral to our understanding of social group dynamics is homophily \cite{mcpherson2001}. A study of the homophily of student groups on a university campus was presented in \cite{flamino2021creation}. It included modeling the evolution of these groups by tracing over time the opinions held by these students on a variety of issues. We found that the most stable groups in terms of stability and longevity of members consisted of students with majority opinions. In contrast, groups with students holding minority opinions were unstable, often changing members and dissolving. We also showed that the entire system evolves toward a stable state in which all groups are fully polarized on the opinions most important to the members. 

The question thus arises of whether the homophily principle and its impact on group dynamics can be observed on online social networks as well. Social networks do not facilitate only interactions between actors, but influence user decisions by content recommendations biased by preference tracking algorithms, such as used by Twitter and other social media. Such preferences are also used by socially-aware networks in which edges represent voluntary social interactions between users and which provide network services using social network analysis techniques \cite{Tsugawa2018}. Furthermore, preferences apply to users interacting with social media features, as some forms of interaction and information consumption are used more than others \cite{Velichety2013}.

Major social media platforms like Facebook, Twitter, and Instagram continue to grow, but such growth is not limited to these highly popular platforms. In fact, recent events in U.S. politics have prompted an entrance of new, alternative platforms to cater to specific groups of users. The most visible example is Parler \cite{mercerwsj}, launched in 2018. This microblogging platform marketed itself as the ``free-speech'' social media alternative. Designed as a Twitter clone, Parler aimed to become a platform for Right Leaning social media users alienated by Twitter. In this paper, we analyze the dynamics of group evolution on social media using data collected by tracking users on Parler and Twitter and assigning them initial and final political biases. They are defined by the average biases of URL links posted by these users during the first and last month of activity, respectively. The tracking of users lasted from September to December 2020, a period that includes the 2020 U.S. Presidential election, which occurred in November of that year and triggered a high level of political interactions during that time.

Using the biases assigned to users, at the end of each period, we created groups of users of the same bias and two ``constellations'' of groups of Left and Right biases, each regardless of the intensity of the respective biases. Then, we analyzed the evolution of these groups on Parler and Twitter. Our analyses confirmed that side-effects of homophily uncovered in our previous work on interactions of students (which ranged from face-to-face meetings to cell-phone messages and calls) are also valid on social media. The two methods of avoiding interactions in diverse groups are either changing important opinions to majority ones, or if this fails, dropping off the platform. Overall, we aim to demonstrate that homophily plays a role in group formation, evolution, and retention online by showing that these results hold in an online setting for two contrasting social media platforms.

Our results show that Twitter has two stable bias groups with the locally largest fractions of members across the political spectrum: liberal bias and conservative bias. They have the local maxima in terms of political bias stability, with holders of these biases retaining their opinions for a long time. In contrast, groups with members holding unpopular political biases were unstable, with their members quitting the platform or leaving to groups with more popular political biases. The desire to interact with peers with similar views motivates holders of unpopular political biases to change their biases or keep them and leave the social media platform. This desire drives the evolution of the large platforms, like Twitter, toward bimodal polarization. In contrast, the smaller platform, Parler, has been dominated from its start by Extreme Right bias and fake news bias, which heavily overlap in terms of committed users making their groups stable and popular. Stability of dominating biases and initially the lack of a noticeable presence of liberal biased content on the platform freezes these two patterns into permanence. The resulting homogeneity formed an unopposed echo chamber on Parler, where the users in this echo chamber engage in and proliferate the same kind of content with little deviation.

\section{Terminology}
While Twitter is an established social media platform that has been subject to numerous research studies across multiple disciplines, Parler has seen less attention. Subsequently, we highlight below the content terminology used within Parler's user interface for those not familiar with the platform.

Parler is fashioned after Twitter, and their methods for content generation and interaction are similar, with different names. Posts on Parler are called ``Parleys''. Parler users are allowed to make posts which are visible to other users and are limited to a maximum of 1,000 characters. Each post can be upvoted or downvoted to indicate if the voting user agrees or disagrees with the content. However, posts only show the number of upvotes, and not the number of downvotes. Comments can be made under posts, and these comments can be upvoted or downvoted. Comments can be made under existing comments on posts, creating a local comment tree. Parler's version of Twitter's retweet feature is the ``echo''. Echoing allows users to choose an existing post and post it to their page, optionally adding content that appears above the post. Users can post a variety of content, including URLs and multimedia. 

To summarize, Parleys are equivalent to Tweets, Echoes to Retweets, and Upvotes to Likes. Comments are similar across Parler and Twitter. 
The equivalence of these features and the intentional similarities between Twitter and Parler means that the graph structure that is organically created by the usage of both websites ends up looking similar too.

\section{Methods}
\subsection{Datasets}
The Parler database was accessed in 2021 \cite{dataarchive}. The published dataset includes most of the posts sent between March of 2018 and January of 2021. It contains about 183 million Parler posts sent by 13 million users. We analyzed a subset of these posts ranging from September $1^{\text{st}}$ to December $1^{\text{st}}$, 2020. The Twitter dataset was obtained from \cite{flamino2021shifting}. This dataset was collected using the Twitter Search API to find all tweets, retweets, quotes, and replies containing the name of one of the two primary 2020 U.S. Presidential candidates sent between June of 2020 and December of 2020. This search yielded approximately 702 million Twitter communications sent by 20 million users. As in the case of Parler, we analyze a subset of all communications sent between September and December 2020. While bots are a problem, analysis of the Twitter dataset shows that only about 1\% of events were generated from unofficial Twitter clients in 2020~\cite{flamino2021shifting}, suggesting that bot presence is limited in our subset. However, as little research has been done on bot detection within Parler, we do not explicitly filter for bots across datasets to maintain consistency and maximize the data available for comparison.

\subsection{News Media Classification}
\begin{figure}[!t]
\centering
\includegraphics[width=2.3in]{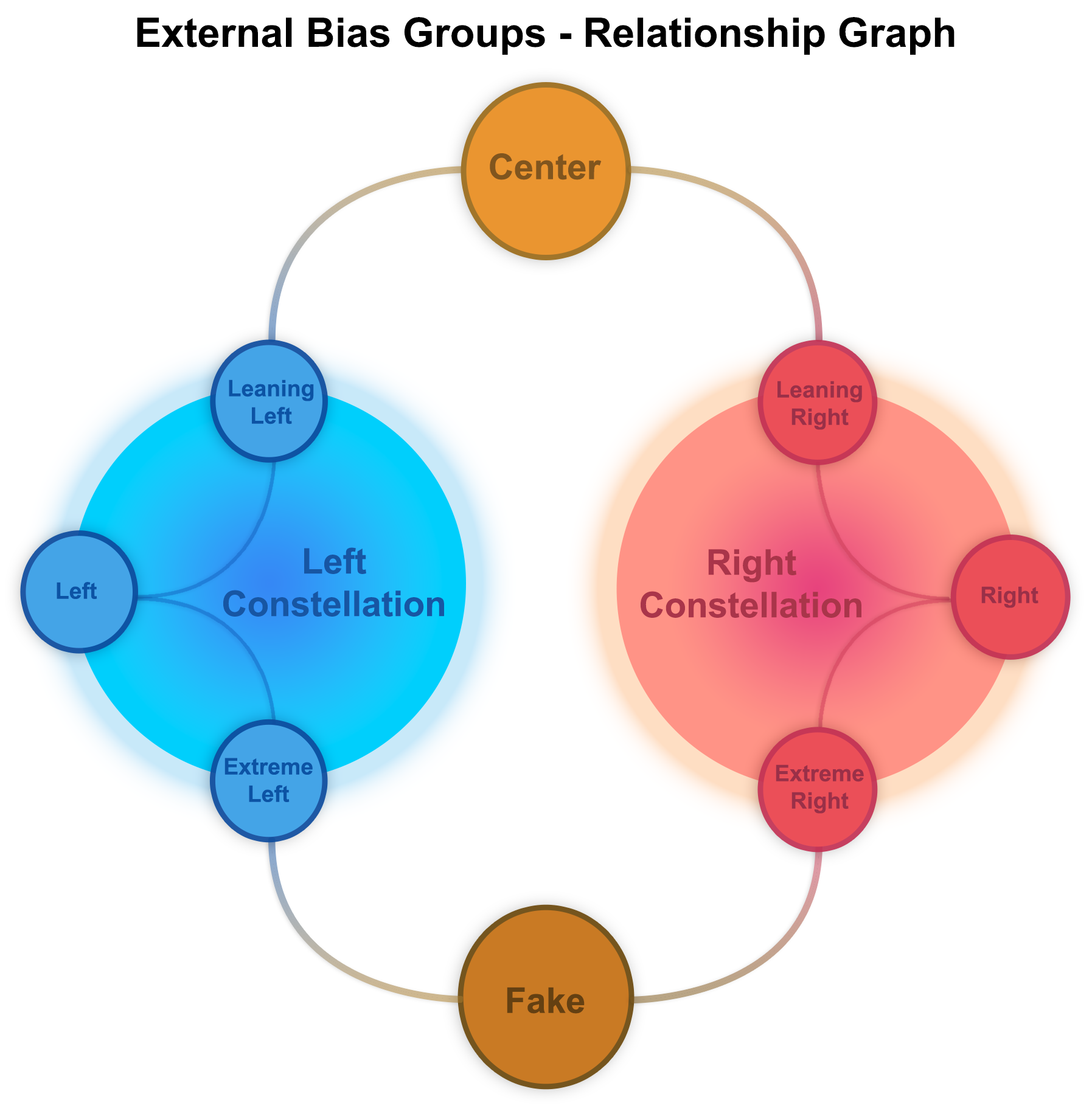}
\caption{A two-level clustering of polarized users. The lower level contains eight bias groups. The higher level consists of two primary clusters called constellations that group associated biases together. The Center bias and Fake news groups exist outside the two constellations. Edges connecting constellation's groups show that members can directly reach groups within each constellation, defining unit distance between them. Travel between groups across the constellations requires several unit steps. Each user has two biases, initial and final. The initial bias uses URL links from the initial month of collected data, while the final bias uses links gathered in the last month. Each user with two different biases travels from initial to final bias, changing the sizes of bias groups dynamically.}
\label{bias_struct_fig}
\end{figure}

We focus our study on the political biases of users on Parler and Twitter. Accordingly, we need to identify the political leanings of the content they propagate. To do this, we adopted a methodology used in \cite{flamino2021shifting}, which was originally designed for Twitter. These classifications identify political biases or fakeness of news media outlets. So, given a Tweet with a URL linking to a valid outlet, we can assign political bias to this tweet. The classifications we use to identify biases of users originated from two 
websites: allsides.com (AS) and Media Bias / Fact Check (MBFC).

AS is a well-known, respected tool for rating news media bias that combines methods like blind surveys, academic research, community feedback, independent reviews and editorial reviews (www.allsides.com/media-bias/media-bias-rating-methods). MBFC assigns news media biases, using a different approach that relies on evaluation of wording, sourcing, story choices, and political endorsement (www.mediabiasfactcheck.com/methodology). MBFC results have been used for labeling bias and factual accuracy of news sources in several academic studies and journal publications. 

The combined evaluations of AS and MBFC have been used to classify a total of 119 media news outlets. The classifications are grouped into five news media categories based on the traditional U.S. political spectrum. Given that the ``Left'' represents liberals and ``Right'' represents conservatives, the categories are {\it Right, Leaning Right, Center, Leaning Left} and {\it Left}. These categories are refined by the addition of two more categories, {\it Extreme Right bias} and {\it Extreme Left bias}. These two categories include news media organizations that tend to exhibit heavy bias toward selected political issues, to the point of promoting propaganda or conspiracy theories not supported by any credible sources. Finally, the third addition, a {\it fake news category}, includes any news media organizations that have been flagged by AS and MBFC as sites that regularly disseminate controversial or fake news to force their points of view. Once these categories are assigned to the news sources, all users can be classified by the content that they consume or spread.

\begin{table*}[t]
    \centering
    \renewcommand{\arraystretch}{1} 
    \caption{The count of news URL links posted on Twitter and Parler, grouped by political bias. The rank column shows numeric values assigned to them. The percentages show the content fractions that fall within that news media category. The average bias of Twitter users is $3.96$ which is Center bias, while for Parler it is $6.83$, Extreme Right bias. Left biases are ranked 1–3, right biases are ranked 5–7, and the center is 4.}
    \begin{tabular}{|*{6}{c|}}
        \hline
        \rowcolor{white} \textbf{News Media Category Bias} & \textbf{Rank} & \textbf{Twitter Count} & \textbf{\% of Twitter Total} & \textbf{Parler Count} & \textbf{\% of Parler Total} \\
        \hline
        \rowcolor{brown} Fake news & 8 & 4,348,747 & 5.96 & 280,502 & 42.30 \\
        \rowcolor{pink} Extreme right & 7 & 4,064,820 & 5.57 & 104,159 & 15.70 \\
        \rowcolor{pink} Right & 6 & 8,691,901 & 11.91 & 199,320 & 30.06 \\
        \rowcolor{pink} Leaning right & 5 & 4,648,000 & 6.37 & 53,402 & 8.05 \\
        \rowcolor{brown} Center & 4 & 7,568,472 & 10.37 & 18,149 & 2.73 \\
        \rowcolor{blue} Leaning left & 3 & 33,093,257 & 45.35 & 5,915 & 0.89 \\
        \rowcolor{blue} Left & 2 & 10,513,306 & 14.41 & 1,504 & 0.22 \\
        \rowcolor{blue} Extreme left & 1 & 39,857 & 0.05 & 167 & 0.02 \\
        \hline
    \end{tabular}

    \label{tab:urls} 
\end{table*}

\begin{figure*}[b]
\centering
\includegraphics[width=5.5in]{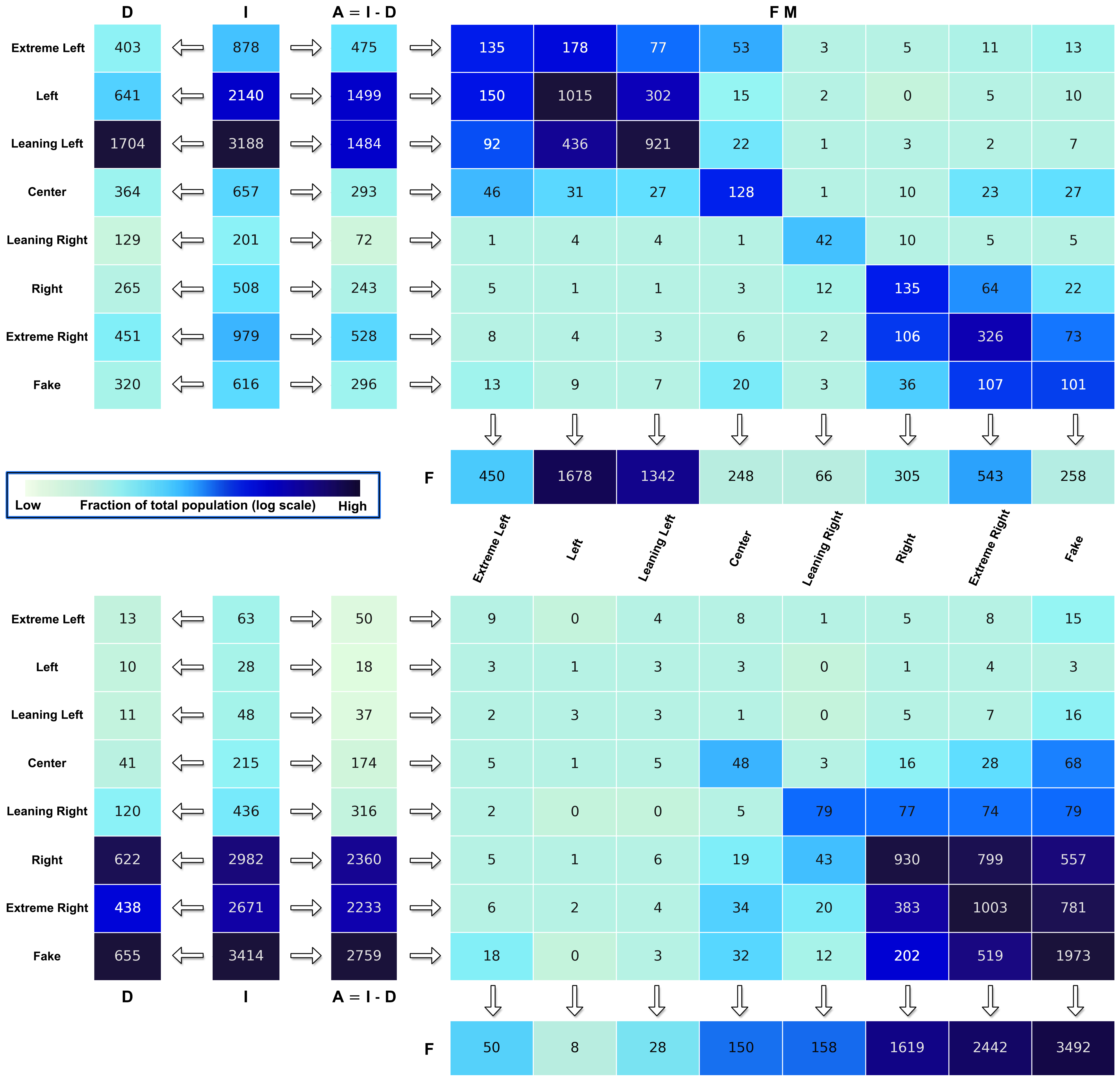}
\caption{Flow diagram of Twitter (Top) and Parler (Bottom) users. Column \textbf{I} shows the number of newcomers in each of the initial bias groups. Column \textbf{D} to the left of \textbf{I} shows the number of newcomers that drop out from the platform. Column \textbf{A} shows the number of newcomers who obtain a final bias classification. The Flow Matrix \textbf{FM} connects active users with the same initial bias to the final bias assigned to them. The bottom row \textbf{F} shows the number of users with their final biases. Thus, the direction of flow is from column \textbf{I} to \textbf{A}, then to columns \textbf{FM} along the corresponding row, and finally to row \textbf{F}.}
\label{FM}
\end{figure*}

\subsection{Mapping Users to Political Bias Groups}
We note that political bias, in the context of social media graph data, is an external characteristic. Therefore, grouping users by their political bias is an external grouping. As opposed to an internal characteristic such as centrality, political bias is a property we ascribe to users based on the political bias classifications of their posts based on AS/MBFC assessments. As such, the classifications of political views and related conclusions contained in this paper should not be interpreted as representing the opinions of the authors or their funders. 

MBFC/AS classification ranges from Extreme Left to Center and Fake News, defining eight classes that we ranked as follows. The Extreme Left is ranked 1, and the remaining classes are listed in the order of increasing Right bias and assigned the rank by 1 larger than its left predecessor, ending with rank 8 assigned to fake news. These URLs ranks are averaged over all URL links posted by this user over the initial and final month of data collection, respectively to obtain the initial and final biases of this user. A user with the initial bias who did not have any posts within the last month is classified as a platform dropout. This method can measure user bias evolution over time using different time periods and define more than two time intervals such as an initial and final month used in this paper, enabling monitoring user bias evolution more precisely.

Fig. \ref{bias_struct_fig} shows two-level clustering of polarized users. At the lower level, there are eight bias groups we defined earlier. 
For group membership, each user rounds its bias to the nearest integer value and joins the corresponding group. At the higher level, we cluster together Left and Right biases regardless of their intensity, which creates Left and Right constellations, with the Center bias and fake news groups existing outside of these constellations. The connections between these groups show that some groups are ideologically ``near'', such as Leaning Left and Left groups, whereas other groups are ideologically ``far''. 
Thus, groups that have a single edge between them are at a unit distance away from each other and would require a member to shift their beliefs a little to move between them. For a pair of groups not connected by an edge, the member of the initial bias group can travel along the shortest path from it to reach the final bias group. The number of edges in that path will define the distance traveled by this member.

Table \ref{tab:urls} displays the total number of classified content items in our Twitter and Parler datasets, grouped by their assigned news media category determined by AS and MBFC. This gives us an initial perspective on the political leanings of these platforms. Parler has a strong conservative news media presence. 
In contrast, Twitter users have more balanced news media usage.

\subsection{Dynamics of User Flows between Bias Groups and Platforms}

We portray the movements of the number of users between political biases over time using a Flow Matrix (FM) in which each row and each column represents a bias group. Each cell in the Flow Matrix shows the number of users that moved from the initial bias of their row to the final bias of their column. However, a decrease in membership numbers occurs between the initial and final population of users in the study. This is because some users stop posting early on and do not post again. We call these users ``dropouts''. We categorize any user who stops posting and does not make a single post for two months or more as a dropout from their platform instead of assigning them a final bias. Thus, these dropouts are not included in the flow matrix calculation.

Using the Flow Matrices, we can find the distance and direction each user moved based on their initial bias group and final bias group. To clarify notation, we assume that clockwise movements in Fig. \ref{bias_struct_fig} and leftward movements in the FM have negative polarization. The corresponding counterclockwise and rightward movements have positive polarization. We calculate these movement vectors for every user and compute the mean, median and Interquartile Range (IQR) of the movement vectors for each bias group. We visualize this data using box plots in the below section, which illustrate the dynamics of inter-group movements for the two platforms.

\section{Results}

\subsection{Dynamics of User Political Biases}

We compute and display the dynamics of users' biases for Twitter and Parler in Fig. \ref{FM}. The new users arrive at the input column labeled ``I'' and each cell of this column represents the number of newcomers with the label of this cell. New users who do not stay long enough to be assigned a final bias flow to the dropout column ``D''. Each cell of column ``D'' has the count of dropouts for each bias category. The remaining newcomers move to the active users column ``A'' to the right of the column ``I''. From there, users leave the cells from the ``I'' vector defining their initial bias to the Flow Matrix ``FM'' in the same rows as their cell and to the column in FM that represents their final bias. Therefore, summing FM along each row yields the number of users with the initial bias represented by this row (this number is stored in vector ``I''). Summing this matrix along the columns yields the number of users whose final bias is represented by their column. These numbers are shown in the bottom row ``F'' and arrows indicate which column shows the composition of initial biases in each final bias cell in column ``F''. 

Fig. \ref{FM} exposes patterns of political bias propagation on Twitter and Parler, revealing an interesting trend in user groupings in each news media category. In Twitter, there are two disjoint communities that have two of the locally largest fractions of users. One community is centered around the Left (liberal) news media category and the other is centered on the Right (conservative) and Leaning Right news media category, with little overlap with the center news media category. In contrast, Parler's FM yields a singular community with a locally largest fraction of users. It is centered around the fake news and Leaning Right bias news media categories. The bimodal and unimodal patterns of Twitter and Parler, respectively, characterize the diversity of news propagated on these platforms. The act of dropping out from a platform can arise in many kinds of human interactions, but with different intensities, as seen in Fig. \ref{FM}.

These figures display the raw numbers for the initial, final, and dropout populations for each bias group on both platforms from which computed fractions of the dropouts in each bias category and observed different dropout trends for different groups. 49.7\% of all users on Twitter dropped out from the platform between September and December, compared to 19.4\% for Parler. This significant difference in dropout fractions highlights the stability of Parler.

In both platforms, the differences between the Right and Left biases were small, a bit over 10\% of the dropout rate in each case. The dropout rate was higher for the Right bias (50.6\%) than the Left bias (45.3\%) on Twitter, but lower for the Right bias (19.2\%) than the Left bias (21.9\%) on Parler. This demonstrates that the existence of only one popular political bias on Parler prevents individuals with biases distant to the popular political bias from even attempting to join Parler, since those who join have a similar rate of staying on the platform as the rate of the user with popular biases. Subsequently, a perpetual echo chamber arises through the overall avoidance of the platform, not from more intensive user dropout.

Comparing these dropout rates to university students \cite{flamino2021creation} reveals that resistance to dropping out is strong in this offline setting, since the yearly dropout rate from the target campus, Notre Dame University, was 2\% (80 to 100 times lower compared to our social media platforms, considering the four-times longer time over which student dropouts were counted). This difference highlights the notably lower cost of dropping out of social media platforms, which can be done in a short time without jeopardizing any long-term relationships. In contrast, students must invest one year of their time before leaving, and usually will have some new acquaintances on campus by that time. They will also likely be subsequently entering a new university with already established groups of students, which can make socialization more difficult. 
 
\subsection{Movement Dynamics}

\begin{figure*}[t]
\centering
\includegraphics[width=6in]{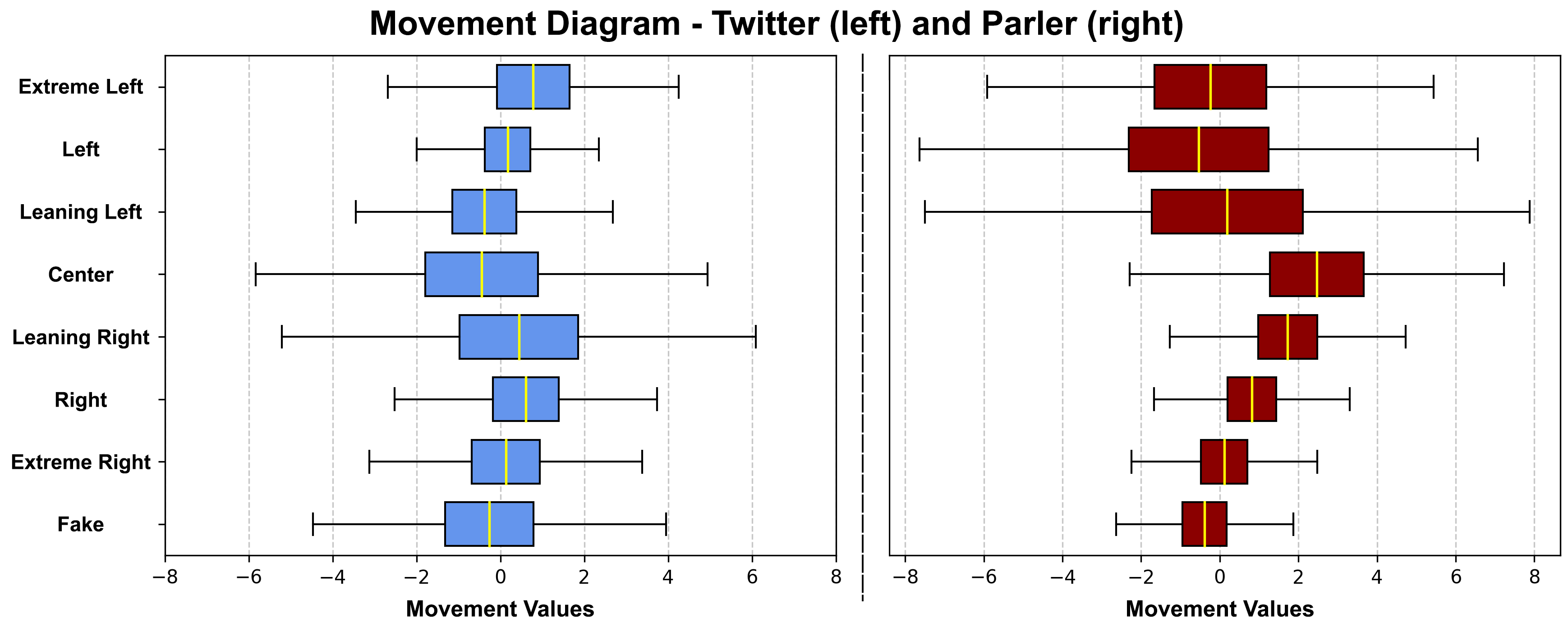}
\caption{Diagram visualizing the movements for each bias group on Twitter and Parler. Each box plot shows the Interquartile Range for the initial-to-final bias group distances traveled by each user initially at that group. The yellow lines in the box plots represent the median distance traveled by the group members and whiskers on either side visualize the maximum extent of distance moved.}
\label{movement_dynamics}
\end{figure*}

The movement diagram for each bias group (Fig. \ref{movement_dynamics}) visualizes the IQR (Interquartile Range) of movement data measured as the number of steps made by each member starting at that group. Each box in the diagram represents the middle half of the data spanning the range from the first to third quartile of movements of group members. The yellow midline represents the second quartile, while whiskers capture the maximum distances traveled by members. As before, the smallest distance is a unit step in each direction, representing one hop over an edge in Fig. \ref{bias_struct_fig}. E.g., the median value of Parler's Center group is about two steps toward the right direction (as opposed to negative two steps, which are moving toward the left).
 
On Twitter, we observe that an average user's movement was within the two closest groups from their initial group because each box plot is within the range from negative two to two steps. The medians are between zero and one step for each group, indicating low intra-group distances and strong polarization between the two constellations. The box plots create a wave-like pattern because these group movements are self-reinforcing. For example, The Leaning Left bias group is one step from the Center bias group which in turn tends to move further left feeding into the Left group, which itself favors unit rightward movement back into the Leaning Left group.

For Parler, the box plots show consistent rightward movements from the Left bias group toward the Center and Leaning Right bias groups with median movement of two steps, since most movements are limited to the range from one to three steps. Very few movements begin within the Right, Extreme Right or Fake bias groups. The Fake and Extreme Right bias groups interact mostly internally leading to the formation of the echo chamber. Parler's Left constellation also shows instability, with a large fraction of users abandoning the platform. These patterns are very different from Twitter's, which neither exhibit a strong directional preference nor constellation-wide instabilities for either Left or Right sides.


\section{Discussion}

In this paper, we collected Parler and Twitter data around the 2020 U.S. Presidential election to compare the political content propagation dynamics of these platforms. This comparison demonstrates fundamental differences in the populations of the two online social mediums. Parler was created to provide an alternative to Twitter, with an emphasis on political free speech attempting to attract alienated users from other social media in the wake of the political discourse triggered by the 2020 election. To provide insight into the type of users that Parler attracted, and the political information being disseminated in Parler and Twitter, we used political bias classifications of news media outlets to identify the presence of fake news and classify content along the U.S. political spectrum. We then characterized the dynamics of content propagation by users by analyzing user movement behavior. These results, combined with our political categorizations of posts on Twitter and Parler, allowed us to show how stable each type of political bias is measured by consistency with which users continue to propagate the content of their current bias. 

On Twitter, we found two consistent and disjoint groups of overlapping users, where liberal-oriented users tended to spread only similarly liberal biased news, while conservative-oriented users spread only similarly conservative biased news, creating two locally largest fractions of the group members. In contrast, Parler had only one distinct group with the locally largest fraction of the group members, lacking any significant patterns of liberal biased news spread. Instead, there were primarily only conservative-oriented users who consistently spread conservative bias and fake news.

Characterizing these patterns, we observed that on Parler the fake news category had the greatest fraction of users migrating to it or choosing to stay in it. This indicated that users on Parler who initially spread fake news had a penchant to continue disseminating them. Furthermore, users with other political biases were more likely to shift and propagate fake news themselves, suggesting the presence of a strong echo chamber. The fake news group on Twitter, on the other hand, did not attract a significant proportion of the members. Instead, Twitter had two bias groups with locally largest fractions of members: one centered around liberal news media categories, and the other centered about conservative news media categories. Subsequently, users with these biases were most inclined to retain them, with similar political biases being likely to migrate to them, causing polarization as users converge on these opposing political biases. We note that the bimodal pattern of Twitter here corroborates observed polarization between the Left biased and Right biased users reported in~\cite{flamino2021shifting}, which also showed a decreasing overlap in center-biased discourse over time.

The broader impact of the results of this paper is the advancement of our understanding of how human behavior adapts to new ways of interpersonal interactions, and how new technologies can benefit from such patterns. One example of this persistence are the trends observed on Twitter and Parler that expand on the results from \cite{flamino2021creation}, which show that university student groups whose members were mostly of majority opinion holders had more stable membership and persisted longer than groups whose members held minority opinions. Parler initially gained majority of fake news and Extreme Right bias and then maintained these biases over time. Meanwhile all liberal biased content was relegated to an insignificant minority. However, on Twitter, users with a broad range of political biases were initially joining, resulting in the formation of two groups of biases. In both cases, the users' behaviors show two tendencies, one for moving toward stable opinions, and another dropping out of the platform. These tendencies drive polarization, as users migrate to stable popular political bias groups, and unpopular outlier biases are deserted, resulting in the formation of isolated echo chambers.

Studies of temporal social networks~\cite{Kulisiewicz2018entropy} show quantitatively that people do not communicate randomly in all types of interactions, which causes entropy of the interactions to decrease over time. The same conclusion is reached in our research, as dynamics of political biases in social media tend to stabilize user interactions over time. Within this scope, we can conclude that optimizing social media and socially aware networks implementations for such patterns \cite{Tsugawa2018} will be efficient.

\section{Future Directions}

The results presented in this paper offer interesting avenues for future work. Among them, graph-based comparisons between Parler and Twitter will likely provide further insights into their differences. Comparing the content characteristics and propagation habits of users on both platforms is of interest, to see if strong content moderation on Twitter led to more accountable behavior from its most influential members compared to Parler. Additionally, integrating agent-based modeling of opinion networks~\cite{sobkowicz2012} with the behavior we observed here can allow for further scaling of bias dynamics beyond the limitations of our current dataset.

We plan to study bias dynamics over time periods smaller than three months. Computing biases periodically on a weekly basis will reveal trajectories over the graph shown in Fig. \ref{bias_struct_fig}. Having them will allow us to measure the forces that (1) attract users to popular bias groups, (2) restrict the length of travel in search of peers, and (3) motivate users to drop out from the current social media platform. The first force is a side-effect of homophily \cite{mcpherson2001}, which is the tendency to interact with people with compatible views. It is easy to ensure such compatibility in small groups of face-to-faces interacting people, yet difficult for technology enabled large interacting groups of social media users. Homophily motivates people to change their views to interact comfortably within such groups. The second force, confirmation bias \cite{Nickerson1998confirmation}, prompts users to choose familiar or similar opinions, constraining the strength of homophily. If the second force prevails, and no close-by stable group exists, the third force, also rooted in homophily, motivates users to leave the social media platform that are incompatible or hostile to the user's views. The second force strengthens with time as long as the biases persist. But the interplay is subtle. When confirmation bias breaks and frees the user to move farther across biases, the user adapts a new bias and confirmation bias switches to it. Thus, after new biases are accepted, they are enforced by confirmation bias and homophily, making new members of a stable group more committed to it than the old ones. We plan to extend this work by adding quantitative analyses of these interesting observations, utilizing measures of utility of membership in groups as seen in~\cite{flamino2021creation}.

For developers of socially aware networks systems, the knowledge of the patterns arising in interactions of users of social media is important. Patterns such as stable and popular groups of users that define social network communities, echo chambers and patterns of real-time data access are essential for designing socially aware caching \cite{Tsugawa2018}. Hence, they can be used for social-based community detection, routing, and data caching strategies and algorithms in social media and social aware networks.

\textbf{Acknowledgements:} This work was partially supported under the DARPA contract HR001121C0165, and the National Science Foundation Grant NSF-SBE 2214216.

\bibliographystyle{unsrt}
\bibliography{bibliography}

\section{Biographies}
Boleslaw K. Szymanski [LM] is the Claire and Roland Schmitt Distinguished Professor of Computer Science and Founding Director of Network Science and Technology Center at RPI. He received the British Computer Society Wilkes Medal, and William H. Wiley 1866 Distinguished Faculty Award, RPI. Since 2011, he is a member of National Academy of Science in Poland. His current interest focus on Network Science and Computer Networks.
\newline

James Flamino [M] received his Ph.D. degree in Physics at RPI in 2021. He is currently a Postdoctoral Researcher at RPI's Network Science and Technology Center. His research applies statistical physics and AI to social systems.
\newline

M. Shahid Modi [SM] is a CS Ph.D. Scholar at RPI's NeST Center. He obtained his B.Tech at Dr. MGR ERI, India. His research uses AI and CS tools to understand social networks.
\end{document}